\documentclass[reprint, prl, superscriptaddress, nobibnotes, amsmath, amssymb, aps]{revtex4-2}

\usepackage{graphicx}
\usepackage{dcolumn}
\usepackage{newtxtext,newtxmath}
\usepackage{bm}
\usepackage[version=4]{mhchem}
\usepackage{amsmath,amssymb}
\usepackage[normalem]{ulem}
\usepackage{braket}
\usepackage{dsfont}
\usepackage{bm}
\usepackage{color}
\usepackage{upgreek}
\usepackage{multirow}
\usepackage[dvipsnames]{xcolor}
\usepackage{svg}
\usepackage{textcomp}

\begin{document}


\title{\textbf{Isolating Exciton Dissociation Pathways  in \ce{ReSe_2} } 
}%

\author{Bradley G. Guislain}
\affiliation{Department of Physics \& Astronomy, University of British Columbia, Vancouver, British Columbia, V6T 1Z1 Canada}
\affiliation{Quantum Matter Institute, University of British Columbia, Vancouver, British Columbia, V6T 1Z4 Canada}


\author{Rysa Greenwood}
\affiliation{Department of Physics \& Astronomy, University of British Columbia, Vancouver, British Columbia, V6T 1Z1 Canada}
\affiliation{Quantum Matter Institute, University of British Columbia, Vancouver, British Columbia, V6T 1Z4 Canada}

\author{Matteo Michiardi}
\affiliation{Department of Physics \& Astronomy, University of British Columbia, Vancouver, British Columbia, V6T 1Z1 Canada}
\affiliation{Quantum Matter Institute, University of British Columbia, Vancouver, British Columbia, V6T 1Z4 Canada}

\author{Giorgio Levy}
\affiliation{Department of Physics \& Astronomy, University of British Columbia, Vancouver, British Columbia, V6T 1Z1 Canada}
\affiliation{Quantum Matter Institute, University of British Columbia, Vancouver, British Columbia, V6T 1Z4 Canada}

\author{Sergey Zhdanovich}
\affiliation{Department of Physics \& Astronomy, University of British Columbia, Vancouver, British Columbia, V6T 1Z1 Canada}
\affiliation{Quantum Matter Institute, University of British Columbia, Vancouver, British Columbia, V6T 1Z4 Canada}
\affiliation{Canadian Light Source, Saskatoon, Saskatchewan, S7N 2V3, Canada}

\author{Jerry Icban Dadap}
\affiliation{Department of Physics \& Astronomy, University of British Columbia, Vancouver, British Columbia, V6T 1Z1 Canada}
\affiliation{Quantum Matter Institute, University of British Columbia, Vancouver, British Columbia, V6T 1Z4 Canada}

\author{Sydney K.Y. Dufresne}
\affiliation{Department of Physics \& Astronomy, University of British Columbia, Vancouver, British Columbia, V6T 1Z1 Canada}
\affiliation{Quantum Matter Institute, University of British Columbia, Vancouver, British Columbia, V6T 1Z4 Canada}

\author{Arthur K. Mills}
\affiliation{Department of Physics \& Astronomy, University of British Columbia, Vancouver, British Columbia, V6T 1Z1 Canada}
\affiliation{Quantum Matter Institute, University of British Columbia, Vancouver, British Columbia, V6T 1Z4 Canada}


\author{Dario Armanno}
\affiliation{Advanced Laser Light Source, Institut National de la Recherche Scientifique, Varennes, Québec, J3X 1P7 Canada}
\affiliation{Department of Physics, Center for the Physics of Materials, McGill University, Montréal, Québec, H3A 2T8 Canada}
\affiliation{Department of Chemistry, McGill University, Montréal, Québec, H3A 2T8 Canada}

\author{Shawn Lapointe}
\affiliation{Advanced Laser Light Source, Institut National de la Recherche Scientifique, Varennes, Québec, J3X 1P7 Canada}

\author{Francesco Goto}
\affiliation{Advanced Laser Light Source, Institut National de la Recherche Scientifique, Varennes, Québec, J3X 1P7 Canada}

\author{Nicolas Gauthier}
\affiliation{Advanced Laser Light Source, Institut National de la Recherche Scientifique, Varennes, Québec, J3X 1P7 Canada}

\author{Fabio Boschini}
\affiliation{Quantum Matter Institute, University of British Columbia, Vancouver, British Columbia, V6T 1Z4 Canada}
\affiliation{Advanced Laser Light Source, Institut National de la Recherche Scientifique, Varennes, Québec, J3X 1P7 Canada}

\author{Andrea Damascelli}
\affiliation{Department of Physics \& Astronomy, University of British Columbia, Vancouver, British Columbia, V6T 1Z1 Canada}
\affiliation{Quantum Matter Institute, University of British Columbia, Vancouver, British Columbia, V6T 1Z4 Canada}

\author{Ziliang Ye}
\affiliation{Department of Physics \& Astronomy, University of British Columbia, Vancouver, British Columbia, V6T 1Z1 Canada}
\affiliation{Quantum Matter Institute, University of British Columbia, Vancouver, British Columbia, V6T 1Z4 Canada}

\author{David J. Jones}
\email{djjones@phas.ubc.ca}
\affiliation{Department of Physics \& Astronomy, University of British Columbia, Vancouver, British Columbia, V6T 1Z1 Canada}
\affiliation{Quantum Matter Institute, University of British Columbia, Vancouver, British Columbia, V6T 1Z4 Canada}

\date{\today}

\begin{abstract}
Strongly bound excitons dominate the optical response in many van der Waals semiconductors, yet distinguishing between the different microscopic processes governing exciton dissociation remains challenging. Using time- and angle-resolved photoemission spectroscopy (TR-ARPES), we independently track exciton and band-edge carrier populations in bulk \ce{ReSe_{2}} under resonant excitation. By studying the fluence dependence and polarization-controlled exciton density dependence of the exciton dissociation process, we distinguish between competing processes and identify exciton photoionization as the microscopic dissociation mechanism. These results establish a population-resolved strategy for resolving exciton-to-carrier conversion pathways in strongly excitonic materials.
\end{abstract}
\maketitle

The semiconducting transition metal dichalcogenides (TMDs) are a family of layered van der Waals materials where the combined effects of quantum confinement and reduced dielectric screening result in exceptionally strong electron-hole Coulomb interactions. As a result, their optical response is dominated by strongly bound neutral excitons \cite{chernikovExcitonBindingEnergy2014}. Once formed, these excitons may scatter into different excitonic states, recombine, or dissociate into excited charge carriers through several distinct microscopic pathways. Each of these dissociation mechanisms results in a distinct dependence of the rate of carrier generation on experimental parameters (incident fluence, etc.), and the rate of carrier generation in turn directly affects key device performance metrics including photoluminescence quantum yield, quantum efficiency for photodetection, and the excitation density limits for excitons. Being able to distinguish experimentally between different exciton dissociation mechanisms is thus crucial for nonlinear optical applications and the design of optoelectronic devices that rely on significant photoconductivity  \cite{satheesh2DRheniumDichalcogenides2023, muellerExcitonPhysicsDevice2018}.

All-optical ultrafast pump-probe techniques have been used extensively to study exciton dissociation processes in semiconductors including: (i) spontaneous dissociation through interactions with defects and interfaces \cite{handaSpontaneousExcitonDissociation2024,perea-causinPhononassistedExcitonDissociation2021,wangFastExcitonAnnihilation2015,wangAugertypeProcessUltrathin2020,karProbingPhotoexcitedCarriers2015,jhaDirectObservationUltrafast2018,allerbeckProbingFreecarrierExciton2021}; (ii) density-driven processes such as exciton-exciton annihilation (EEA) and excitonic Mott transitions \cite{sunObservationRapidExciton2014,yuanExcitonDynamicsAnnihilation2015,kumarExcitonexcitonAnnihilationMoSe22014,engelUltrafastRelaxationExciton2006}; and (iii) optically driven dissociation via excited state absorption (ESA) \cite{mengAnisotropicSaturableExcitedState2018,linNarrowbandHighlyingExcitons2021,silvaEfficientExcitonDissociation2001,zenzFemtosecondPhotocurrentExcitation2001}. Although optical probes lack momentum resolution and typically must infer exciton and free carrier populations indirectly from changes in the optical response, they have enabled enormous progress in understanding exciton dynamics and decay processes. 
Recently, time- and angle-resolved photoemission spectroscopy (TR-ARPES) \cite{boschiniTimeresolvedARPESStudies2024} has been used as a complementary technique to overcome these limitations in the study of excitons in a variety of semiconducting materials \cite{madeoDirectlyVisualizingMomentumforbidden2020b, dongDirectMeasurementKey2021b, manExperimentalMeasurementIntrinsic2021a, reutzelProbingExcitonsTimeresolved2024, volckaertExternalScreeningLifetime2023, gosettiUnveilingExcitonFormation2025,emmerichUltrafastChargeTransferExciton2020a,stadtmullerStrongModificationTransport2019,tanimuraDynamicsIncoherentExciton2019,tanimuraSurfaceExcitonFormation2023, greenwoodExcitonQuenchingGrain2025,karniStructureMoireExciton2022,schmittFormationMoireInterlayer2022a}. TR-ARPES allows independent monitoring of occupied and excited state populations, enabling researchers to disentangle and track the dynamics of electrons, holes, and excitons after optical excitation. 

\begin{figure}
	\centering
	\includegraphics[scale=0.5]{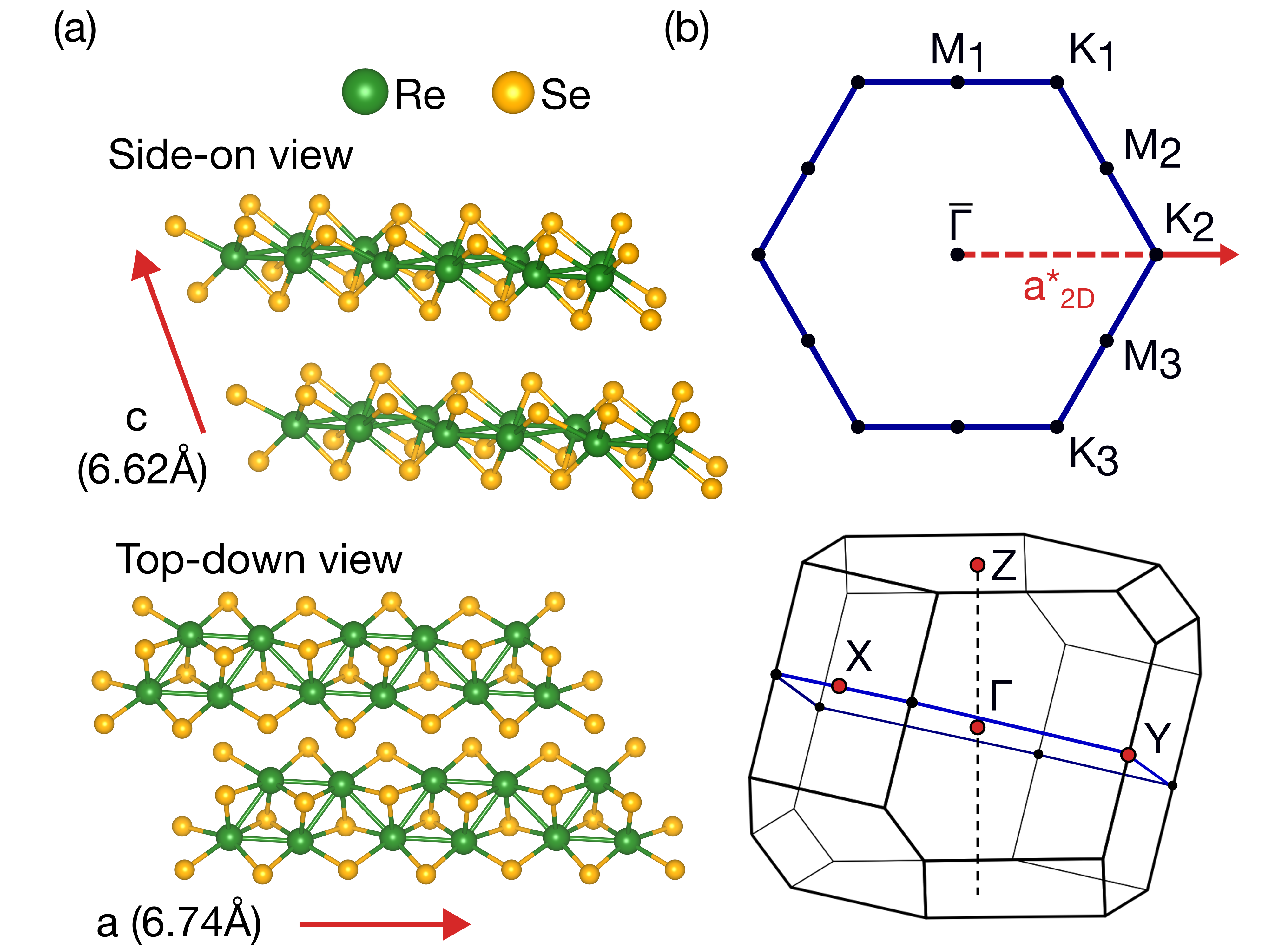}
	\caption {\ce{ReSe_{2}} structure: (a) Real-space crystal structure with lattice constants reported by \textcite{jariwalaSynthesisCharacterizationReS22016} and crystal structure visualized using VESTA \cite{mommaVESTA3Threedimensional2011}. b) Reciprocal-space structure of \ce{ReSe_{2}}. As discussed in the main text, the crystal lattice is triclinic but the 2D-projected Brillouin zone is often approximated as hexagonal with three pairs of inequivalent $K$ and $M$ points. The Re-Re chain axis, along which carriers near the VBM have the smallest effective mass \cite{hartElectronicBandstructureVan2017, kimStrongOneDimensionalCharacteristics2019}, is conventionally taken to be along the $\bar{\Gamma}-{K_{2}}$  direction in reciprocal space.}
	\label{fig:structure}
\end{figure}

Here, we use TR-ARPES to directly monitor spectral weight transfer from excitons to charge carriers in bulk \ce{ReSe_{2}}, a semiconducting member of the TMD family. Unlike the hexagonally symmetric Group VI TMDs, \ce{ReSe_{2}} crystallizes in a low symmetry 1T$^\prime$ triclinic crystal structure, depicted in Fig. \ref{fig:structure}. A key advantage of \ce{ReSe_{2}} for this study is the strongly anisotropic nature of the material's optical response \cite{aroraHighlyAnisotropicInPlane2017, dasAnisotropicExcitonsBulk2025}, which allows us to tune the exciton density following photoexcitation by rotating the pump polarization without changing the incident pump fluence.

Determining the dependence of the free carrier yield on pump fluence as well as the initial exciton density enables us to differentiate between exciton dissociation mechanisms and directly identify that the dominant mechanism for the generation of free carriers is ESA (or exciton photoionization). In this process excitons undergo an optical transition to an unbound electron-hole pair state through the absorption of a second pump photon within the duration of the pump pulse. We model the coupled exciton and charge carrier dynamics using a phenomenological model that reproduces the observed exciton and charge dynamics. In addition, we obtain a photoemission measurement of the bandgap (1.54~eV) and an experimental determination of the excitonic Bohr radius (19.3~$\Angstrom$) in bulk \ce{ReSe_{2}}.

TR-ARPES experiments are performed at the UBC-Moore Center for Ultrafast Quantum Matter \cite{dufresneVersatileLaserbasedApparatus2024} and at the TR-ARPES endstation of the Advanced Laser Light Source (ALLS) user facility \cite{longaTimeresolvedARPESProbe2024}. At UBC, TR-ARPES measurements employ a pump photon energy of 1.55~eV (near-resonant with the band edge transition) and a probe photon energy of 6.2~eV. At ALLS, TR-ARPES measurements use a pump photon energy of 1.38~eV (near-resonant with an excitonic transition) and a probe photon energy of 6.0~eV. At UBC and ALLS, the samples are held at 10~K and 100~K respectively. The choice of temperature is made to avoid the thermal decay of excitons caused by strong exciton-phonon interactions in \ce{ReSe_{2}} above 100~K \cite{dasAnisotropicExcitonsBulk2025}, and to limit sample charging which can differ across samples. In both experiments, the overall temporal and energy resolutions are approximately 350~fs and 20~meV.

\begin{figure*}
	\centering
	\includegraphics[scale=1.0]{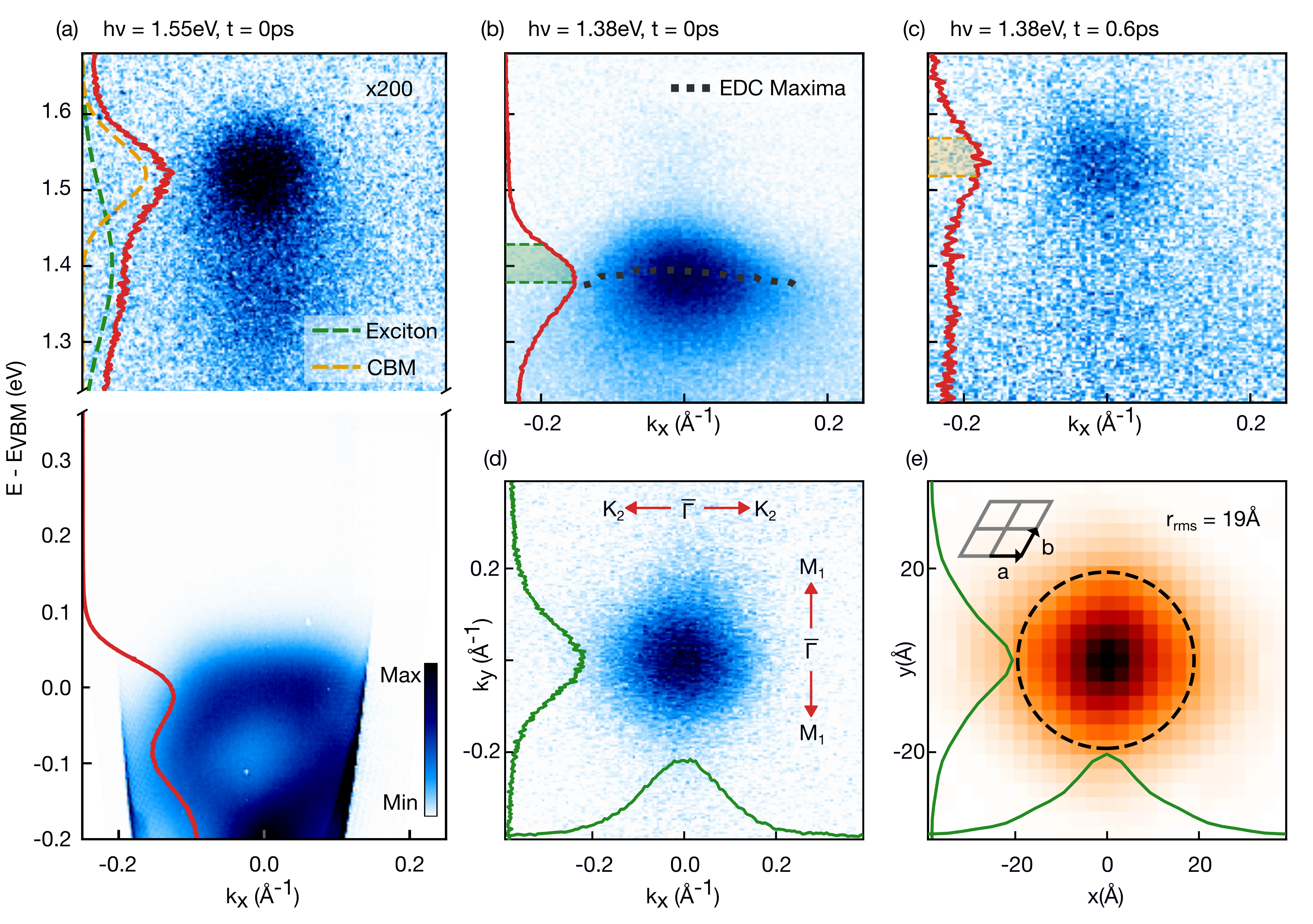}
	\caption{Comparison of the excited state populations in bulk \ce{ReSe_{2}} observed in TR-ARPES experiments using two pump photon energies: (a) near the band edge transition; (b) and (c) near-resonant with the excitonic transition. The image in panel (c) is scaled up by a factor of 6.25 relative to panel (b). Panel (a) displays a cut offset by 24 degrees from the $\bar{\Gamma}$-$K_{2}$ direction, and panels (b) and (c) display a cut along the $\bar{\Gamma}$-$K_{2}$ direction. In the case where the pump is resonant with the exciton, the in-plane component of the pump polarization is parallel to the Re-Re chain direction. Energy axes are referenced to the peak photoemission intensity at the valence band maximum (VBM), and the momentum integrated intensity is shown on red on the vertical axes, along with a two-peak fit in panel (a) representing the CBM and exciton. The pump fluence is 1.6~$\text{mJ}/\text{cm}^{2}$ in the resonant case, and 0.8~$\text{mJ}/\text{cm}^{2}$ in the above-gap case. Shaded regions in (b) and (c) indicate integration ranges for dynamics shown in Fig. \ref{fig:dynamics}. d) Momentum distribution of the exciton photoemission intensity at zero delay and a center energy of $E-E_{VBM}=$ 1.39~eV. (e) Spatial excitonic wavefunction as calculated by 2D Fourier transform of the momentum distribution shown in panel (d), with the unit cell size of \ce{ReSe_{2}} drawn to scale in the inset.}
	\label{fig:energetics}
\end{figure*}   

GW-BSE calculations and supporting optical studies obtain a 1.49~eV bandgap and 120~meV exciton binding energy in bulk \ce{ReSe_{2}} \cite{aroraHighlyAnisotropicInPlane2017}. We pump above the  bandgap at 1.55~eV, and the resulting photoemission spectrum along $K_{2}-\bar{\Gamma}-K_{2}$ is shown in Fig.~\ref{fig:energetics}(a). The photoemission intensity integrated between $\pm$0.1$\Angstrom^{-1}$ is displayed in red on the vertical axis and clearly shows -- in addition to the valence bands below the Fermi level -- a strong peak near the direct transition energy as well as a distinct shoulder at slightly lower energy, both appearing above the Fermi level at zero delay. By fitting the angle integrated intensity to a two-peak model, we find that the separation between the two peaks is 120$\pm$15~meV. Based on the predicted bandgap and exciton binding energy, we assign the higher energy peak to the conduction band minimum (CBM) and the lower energy peak to the exciton. Details of the fitting procedure for the energy distribution curves (EDCs) are shown in Fig. S1 in the Supplemental Material. In this case, our understanding is that excitons are rapidly formed from free carriers on timescales significantly shorter than our time resolution.

In contrast, when we resonantly pump the exciton (below the bandgap) at 1.38~eV, the exciton appears at zero delay as shown in  Fig.~\ref{fig:energetics}(b). In the following 600~fs, the exciton decays while a signal from free carriers at $E-E_{VBM}=$ 1.54~eV rises in intensity as can be seen in Fig.~\ref{fig:energetics}(c). Before discussing the dynamics of this free carrier generation process, we take the opportunity to characterize the exciton. Notably, the excitonic feature exhibits hole-like dispersion as can be seen in the dotted line indicating EDC maxima in Fig. \ref{fig:energetics}(b), consistent with the expected photoemission signature of excitons \cite{rustagiPhotoemissionSignatureExcitons2018a,christiansenTheoryExcitonDynamics2019a}. A constant energy surface ($E-E_{VBM}=$ 1.39~eV) is shown in Fig.~\ref{fig:energetics}(d). The spatial extent of the excitonic wavefunction can be obtained by Fourier transforming this distribution \cite{dongDirectMeasurementKey2021b,reutzelProbingExcitonsTimeresolved2024}, and the results are shown in Fig.~\ref{fig:energetics}(e). We find a root mean square radius of 19.3$\pm$0.6~$\Angstrom$, very similar to excitonic Bohr radii previously obtained for monolayer \ce{ReSe_{2}} on bilayer graphene \cite{volckaertExternalScreeningLifetime2023} and somewhat larger but comparable to the exciton radius for bulk \ce{ReSe_{2}} obtained using GW-BSE calculations \cite{aroraHighlyAnisotropicInPlane2017}. 

\begin{figure}
	\centering
	\includegraphics[scale=0.5]{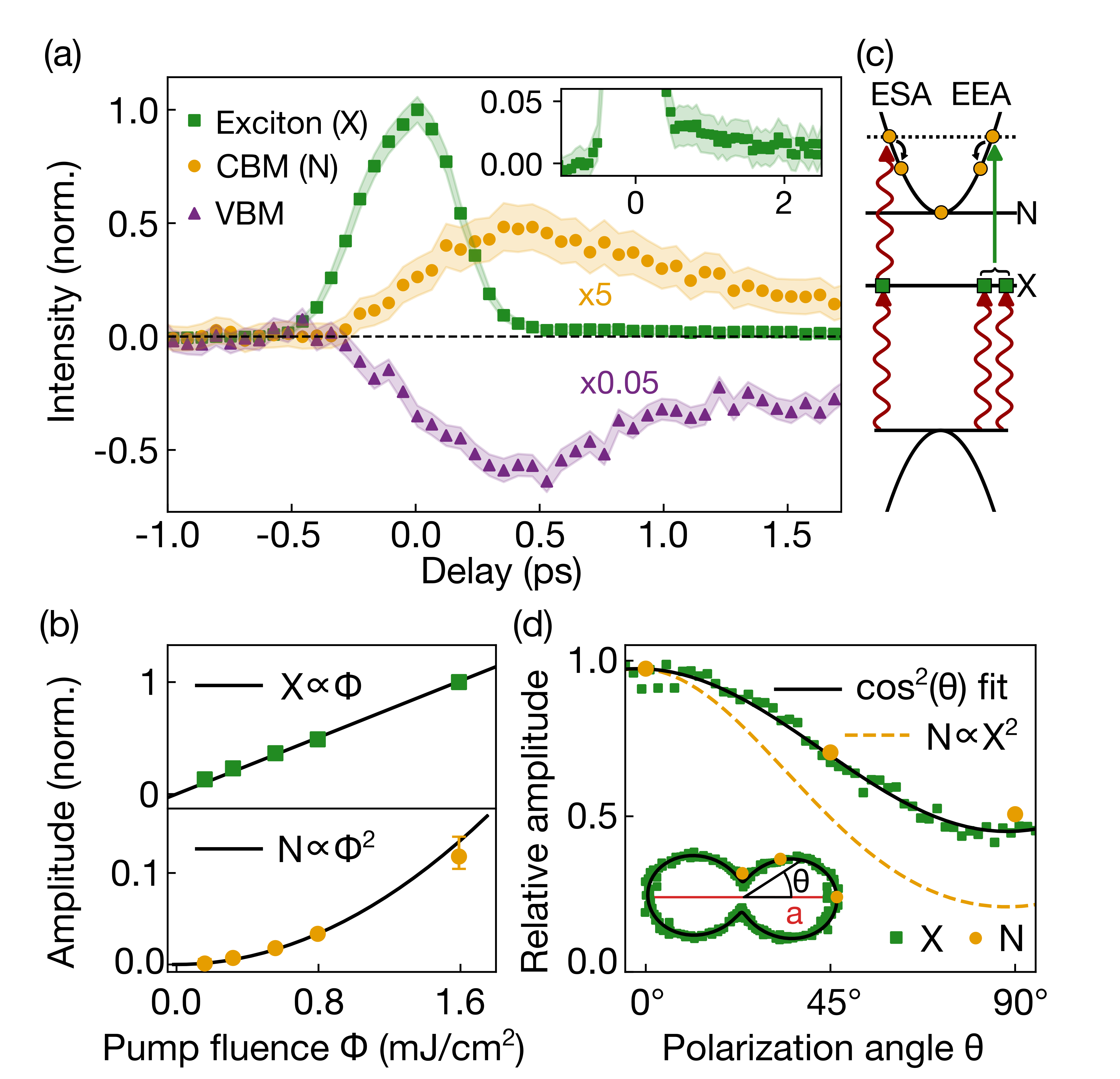}
\caption{Exciton and carrier dynamics when the exciton transition is pumped resonantly. (a) Angle- and energy-integrated dynamics for the exciton, CBM, and VBM populations. See shaded regions in Fig.~\ref{fig:energetics} for energy integration ranges. The inset presents an expanded view of the exciton dynamics to highlight the slowly decaying residual exciton population (same axes scaling). Error bands indicate the standard deviation of each data point. (b) Fluence dependence of the peak exciton and CBM photoemission intensity. Both are normalized to the maximum intensity of photoemission from the exciton for 1.6~mJ/cm$^{2}$ pump fluence. (c) Diagram of two candidate processes for free carrier generation. ESA, in which single excitons are photoionized, is depicted on the left. The right side of the diagram depicts EEA in which two excitons annihilate one another. (d) Linear polarization dependence of the exciton and CBM population with 1.6~mJ/cm$^{2}$ pump fluence, demonstrating that the CBM population does not scale quadratically with exciton population. Each population is individually normalized to its maximum value at $\theta=0^{\circ}$. The inset is the full 360$^{\circ}$ polar plot for reference.}
	\label{fig:dynamics}
\end{figure}

Next, we track the coupled population dynamics of the exciton, CBM electron, and VBM hole populations in the first several picoseconds after resonant excitation of the exciton. The results are shown in Fig.~\ref{fig:dynamics}(a) at a pump fluence of 1.6~mJ/cm$^{2}$. A comparison to the dynamics in the case of above-gap pumping is shown in Fig. S2 of the Supplemental Material. The integration range for the exciton population is chosen to be slightly above the peak photoemission intensity of the exciton to avoid including intensity from pump-probe photoemission from image potential states at lower kinetic energies. The integration range used for for the VBM is shown in Fig. S3 of the Supplemental Material. As mentioned, the exciton population rises promptly at $t_{0}$ and the majority of the population decays on very short ($\leq$100~fs) timescales. Such a decay rate for excitons is much faster than what would normally be expected in the bulk group-VI TMDs, although sub-picosecond decay components for excitons in both few-layer and bulk rhenium TMDs have been reported \cite{volckaertExternalScreeningLifetime2023,huoThicknessdependentExcitonRelaxation2023,chowdhuryRobustCoherentDynamics2024}. Remarkably, the band extrema do not begin to be populated by carriers during the excitation pulse, but rather begin filling on a $\approx 100$-fs timescale and peak roughly 450~fs after zero delay. 

Determining the microscopic mechanism for the generation of free carriers first requires an understanding of how the production of both excitons and free carriers scales with photon flux as shown in Fig.~\ref{fig:dynamics}(b). We find that the exciton and CBM populations fit well to a linear and quadratic dependence on fluence, respectively. Based on the delayed growth of the CBM  population relative to zero time delay, as well as the quadratic dependence of the CBM population on pump fluence, we can conclusively determine that electron-hole pairs are not being generated directly in a one-photon absorption process or through a spontaneous exciton dissociation process involving a single exciton. We can additionally rule out a direct two-photon absorption exciting free carriers, as our pump is resonant with a strong single photon transition. Moreover, previous studies have shown that nonlinear optical behaviour in TMDs at wavelengths resonant with an exciton transition is due to saturable single photon absorption or excited state absorption rather than two-photon absorption \cite{liangSaturableAbsorptionFewlayer2025, mengAnisotropicSaturableExcitedState2018}. We are left with two processes that could feasibly result in the observed free carrier population, both shown schematically in Fig.~\ref{fig:dynamics}(c): (i) ESA (or exciton photoionization), where a previously formed exciton absorbs a second photon to create a free electron-hole pair in a two step process; or (ii) EEA, where two photons create two excitons that subsequently interact to form a free electron-hole pair. While ESA could exhibit a sub-quadratic power-law dependence of the carrier population on pump fluence, in our case the very short exciton lifetime in comparison to the pump duration indicates the power should be very close to 2. In both cases, the finite rise time that we observe for the CBM population should be interpreted as the relaxation time for highly excited carriers decaying to the band minimum.

As both ESA and EEA are processes involving two photons, it is not possible to distinguish between the two based only on observing the pump fluence dependence of the carrier generation rate. To solve this problem, we take advantage of the highly anisotropic response of the exciton transition in \ce{ReSe_{2}}. Figure~\ref{fig:dynamics}(d) shows the polarization dependence of the exciton and CBM populations while incident fluence is held constant. We find that the exciton density is highest when the in-plane component of the pump polarization is near parallel to the crystal $a$-axis, consistent with previous optical studies \cite{aroraHighlyAnisotropicInPlane2017, dasAnisotropicExcitonsBulk2025}. The inset is the standard polar plot, while the main plot displays the unwrapped first quadrant. The polarization dependence of the CBM population for three different polarizations is shown as yellow markers. It is evident that the CBM population scales almost linearly with exciton population rather than quadratically, as would be expected in the case of EEA (the expected quadratic dependence in this case is shown as a dashed yellow line for comparison). Provided that the ionization step has a weak polarization dependence, the linear scaling with exciton population is precisely what would be expected from ESA. Since we are able to exclude EEA based on the observed polarization dependence, we conclude that ESA is the dominant mechanism for free carrier generation in this fluence regime.

\begin{figure}
	\centering
	\includegraphics[scale=0.5]{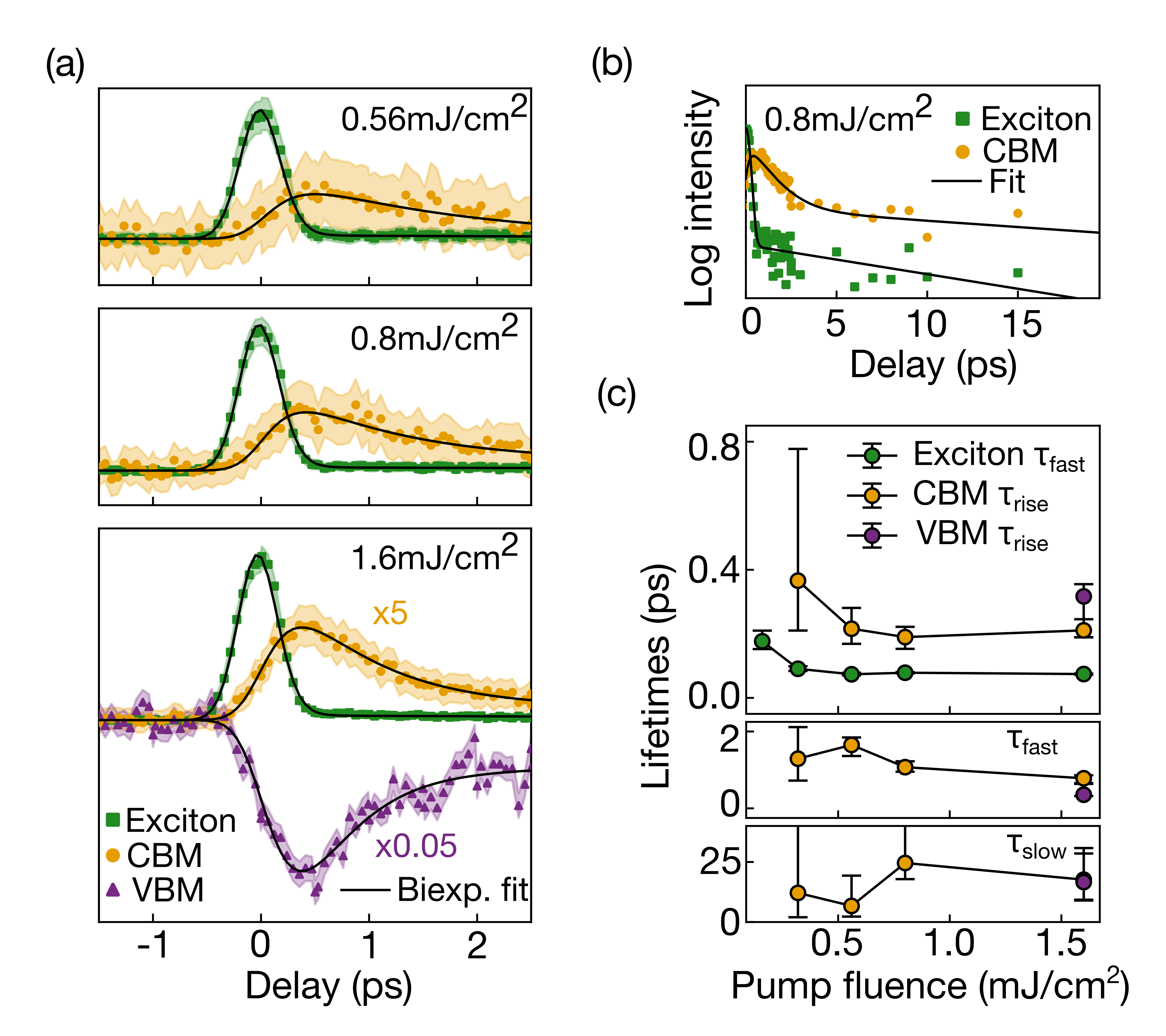}
	\caption{Fit results for the dynamics of the exciton, electron, and hole populations. (a) Best fit results for the biexponential model at the three highest fluences tested over a range of pump-probe delays from -1.5 to 2.5~ps. (b) Dynamics with 0.8~\textmu J/cm$^{2}$ fluence plotted over an extended delay range to illustrate the presence of slower decay channels.
    (c) Best fit lifetimes vs. pump fluence with 86\% bootstrap confidence intervals.}
	\label{fig:fitresults}
\end{figure}

The coupled excitation and decay dynamics of excitons and free carriers are modeled using a standard biexponential model for the excitons, and a phenomenological rate equation model for the free carrier population. The solution to this system of rate equations has the form of a biexponential decay with a rise time $\tau_{\text{rise}}$, which represents the rate that highly excited carriers resulting from ESA relax into the CBM. The fits for the exciton and free carrier populations are not performed independently, but rather as a global fit where the models are forced to share the same $t_{0}$, which means that excitons and highly excited free carriers are populated simultaneously in the model.  A complete description of the model and full report of the fit parameters is available in the Supplemental Material.

The results of fitting to this model for the three highest fluences measured are shown in Fig.~\ref{fig:fitresults}(a). We obtain good agreement for all pump fluences, though at the highest fluences tested it is likely that our instrumental time resolution limits the value that we obtain for the fast component of the exciton lifetime, and the values we report may need to be interpreted as an upper bound. The fast component of the exciton lifetime and the rise time of the CBM population obtained from fitting are approximately 0.2~ps and 0.4~ps respectively at the lowest fluences tested. Both of these timescales decrease by a factor of approximately 2 as the pump fluence is increased by an order of magnitude, and the fast component of the exciton lifetime is significantly shorter than in the case of above gap excitation (discussed further in the Supplemental Material), indicating that the exciton photoionization process plays a role in the exciton dynamics observed. A recent TR-ARPES study of MBE-grown monolayer \ce{ReSe_{2}} on a bilayer graphene substrate \cite{volckaertExternalScreeningLifetime2023} reported exciton decay timescales in the hundreds of femtoseconds and attribute the short lifetime to interaction with defect states and charge transfer into the substrate. We find a similarly short exciton lifetime when we create excitons by pumping above the bandgap in the absence of significant exciton photoionization, though in our case we are measuring a bulk crystal and the fast decay can not be attributed to charge transfer.

The free electron-hole pairs produced by the exciton photoionization process persist for much longer than the primary exciton population, with fast and slow decay timescales of approximately 1~ps and 20~ps. The former is consistent with prior transient absorption studies of bulk \ce{ReSe_{2}} \cite{heUltrafastTransientAbsorption2018}, while the latter is significantly faster compared to previous work. The rise times that we observe for the CBM population are similar to the timescale for the relaxation of hot carriers to the band edge observed recently using TR-PEEM in \ce{6L-ReS_{2}} \cite{qinPolarizationAnisotropyUltrafast2024}, which is consistent with our interpretation of the dynamics. Our TR-ARPES measurements sample only a limited range of $k_{z}$, meaning the features we label CBM and VBM may need to be interpreted as local rather than global band extrema. However, the lifetimes in the tens of picoseconds that we observe for the free carrier populations indicate that the dynamics are not a result of electron-electron interactions as would be expected if we were observing carriers at a $k_{z}$ far from the global band extrema. In the case of the exciton, we expect the $k_{z}$ dispersion to be relatively flat, as GW-BSE calculations \cite{aroraHighlyAnisotropicInPlane2017,dasAnisotropicExcitonsBulk2025} have indicated that the exciton is mainly confined to individual layers of \ce{ReSe_{2}}. 

Using TR-ARPES, we have directly resolved the dissociation dynamics of strongly bound excitons in bulk \ce{ReSe_2} by independently tracking exciton and free carrier populations. Our results isolate exciton photoionization as the dominant dissociation mechanism for excitons in bulk \ce{ReSe_{2}} under resonant excitation, distinguishing it from many-body exciton annihilation processes and explaining the observed sub-picosecond transfer of excitonic population to excited charge carriers at the band-edge. By disentangling exciton decay from carrier thermalization and recombination, this work resolves a prevailing ambiguity in the interpretation of ultrafast optical measurements of excitonic systems. Previous all-optical studies of bulk \ce{ReS_{2}} proposed excited-state absorption as a possible mechanism for anisotropic reverse saturable absorption in the material, but lacked the ability to distinguish and track exciton and carrier populations directly and independently. In addition, our measurements provide the first photoemission based determination of the bandgap and excitonic Bohr radius in bulk \ce{ReSe_{2}}. More broadly, these results establish how population-resolved techniques like TR-ARPES can uncover the microscopic origins of exciton dynamics in semiconductors by unraveling various exciton dissociation pathways from each other, as well as from exciton recombination and carrier relaxation processes.

The authors thank George Sawatzky, Alexander Kemper, and Christina Hofer for helpful discussions, and MengXing Na for assistance with early experimental work. This research was undertaken thanks in part to funding from the Max Planck–UBC–UTokyo Centre for Quantum Materials and the Canada First Research Excellence Fund, Quantum Materials and Future Technologies. This project is also funded by the Natural Sciences and Engineering Research Council of Canada (NSERC), Canada Foundation for Innovation (CFI); the Department of National Defence (DND); the British Columbia Knowledge Development Fund (BCKDF); the Gordon and Betty Moore Foundation’s EPiQS Initiative, Grant GBMF4779 to A.D. and D.J.J.; the Canada Research Chairs Program (A.D.); and the CIFAR Quantum Materials Program (A.D.).

\clearpage
\bibliography{dynamicspaper_bbt}

\end{document}